# Microstructural and transport properties of superconducting FeTe$_{0.65}$Se$_{0.35}$ crystals

A G Sivakov[1], S I Bondarenko[1], A I Prokhvatilov[1], V P Timofeev[1], A S Pokhila[1],

V P Koverya[1], I S Dudar[1], S I Link[1], I V Legchenkova[1], A N Bludov[1], V Yu Monarkha[1],

D J Gawryluk[2,3], J. Pietosa[2], M Berkowski[2], R Diduszko[2], R Puzniak[2] and A Wisniewski[2]

[1] B. Verkin Institute for Low Temperature Physics and Engineering, National Academy of Sciences of Ukraine, pr. Lenina 47, 61103 Kharkov, Ukraine

[2] Institute of Physics, Polish Academy of Sciences, Aleja Lotników 32/46, PL-02668, Warsaw, Poland

[3] Laboratory for Scientific Developments and Novel Materials, Paul Scherrer Institute, CH-5232 Villigen, PSI, Switzerland

E-mail: wisni@ifpan.edu.pl

**Abstract**

The issue concerning the nature and the role of microstructural inhomogeneities in iron chalcogenide superconducting crystals of FeTe$_{0.65}$Se$_{0.35}$ and their correlation with transport properties of this system was addressed. Presented data demonstrate that chemical disorder originating from the kinetics of the crystal growth process significantly influences the superconducting properties of an Fe-Te-Se system. Transport measurements of the transition temperature and critical current density performed for microscopic bridges allow us to deduce the local properties of a superconductor with microstructural inhomogeneities, and significant differences were noted. The variances observed in the local properties were explained as a consequence of weak superconducting links existing in the studied crystals. The results confirm that inhomogeneous spatial distribution of ions and small hexagonal symmetry nanoscale regions with nanoscale phase separation also seem to enhance the superconductivity in this system with respect to the values of the critical current density. Magnetic measurements confirm the conclusions drawn from the transport measurements.

PACS numbers: 74.25.Ha, 74.70.Xa, 74.62.En, 74.62.Dh.

# 1. Introduction

Superconductivity in layered iron (Fe) chalcogenides ("11"-type system compounds) has been the subject of intensive research for the last few years (e.g. [1-4]), performed in order to understand the mechanism of superconductivity, to understand the interplay between superconductivity and microstructure, as well as to increase the critical temperature, $T_c$. Partial substitution of Te for Se leads to an increase of $T_c$ up to about 14 K for $Fe_{1-y}Te_{1-x}Se_x$ with $0.4<x<0.8$ and $y \approx 0$ [5, 6]. The high upper critical field, $H_{c2}$, of this system and the low toxicity of their starting materials compared to the FeAs-based superconductor makes it a promising material for applications in new types of high-field superconducting tapes [7], wires [8], and thin films [9]. Although $FeTe_{1-x}Se_x$, in principle, appears to be an almost ideal model system for the study of the phenomenon of superconductivity in Fe-based compounds, the detailed analysis of the data is significantly hindered by an intrinsic (unavoidable off-stoichiometry driven by the complex phase diagram) and the extrinsic crystal disorder, which results from a complex structural chemistry and an apparent inherent non-stoichiometry. Multi-scale lattice disorder begins at the short range atomic level in the mixed crystal because the Te and Se ions are at slightly different positions in the unit cell [10, 11]. On larger distance scales, crystals of $FeTe_{1-x}Se_x$ tend to have Fe non-stoichiometry often described as the $Fe_7(Te_{1-x}Se_x)_8$, non-homogeneities like clustering, and microstructural foreign phases of Fe chalcogenides [12-16]. Thus, it is a rather difficult task to grow high quality single-phase single crystals of the Fe-Te-Se system.

Recently, we obtained high quality single-phase crystals for the compounds with Se content $x = 0.35$ [17]. The critical temperature $T^{onset}_c$ of the crystals of $FeTe_{0.65}Se_{0.35}$, as deduced from magnetic measurements, was equal to about 12.5 K, and the full width at half-maximum (FWHM) of the rocking curve scan ω of the 004 diffraction peak for the highest crystallographic quality single crystals has been found to be as small as Δω = 1.35 arc min. In our previous papers [17, 18], we studied the morphology of these crystals by transmission electron microscopy and high angle annular dark field scanning transmission electron microscopy. Our data demonstrated a presence of nanometre scale hexagonal-like regions coexisting with a tetragonal host lattice, a chemical disorder demonstrating non-homogeneous distribution of host atoms in the crystal lattice, as well as hundreds-of-nanometres-long Fe-deficient bands. We compared these findings with the magnetic and superconducting properties characterized with magnetization, specific heat, and magnetic resonance spectroscopy. The main conclusion of these studies is that the crystals of apparently inferior quality exhibit more pronounced superconductivity and sharper superconducting transition. In the current paper, we are going to discuss the impact of microstructure on the transport critical current in these crystals.

## 2. Sample preparation

The studied crystals of nominal composition $FeTe_{0.65}Se_{0.35}$ were grown applying Bridgman's method with two different velocities during two procedurally identical growth processes [17]. The samples were prepared from stoichiometric quantities of Fe chips (3N5), tellurium powder (4N) and high purity Se powder (5N). All of the materials were weighed and mixed in an argon filed glove box. Double walled evacuated ($9.32 \cdot 10^{-5}$ Pa) and sealed quartz ampoules with starting materials were placed in a furnace with an average vertical gradient of temperature equal to ~ 0.4 °C/mm and ~ 1.0 °C/mm for the samples described as A and B, respectively. The material was synthesised for 6 h at a temperature up to 700 °C. After melting at ~ 860–880 °C the temperature was held for 3 h, and then was reduced down at a rate of 2 °C/h for sample A and 1 °C/h for sample B. Therefore, the growth velocities of the crystals were equal to ~ 5 and ~ 1 mm/h for samples A and B, respectively.

## 3. Experimental details

Structural and magnetic measurements were performed in order to characterize the studied samples. Optical microscopy, with maximal spatial resolution at a level of 2 μm, was utilised to ascertain the differences in the features of the surface structure of the studied samples. X-ray diffractometry was employed to ascertain the differences in the features of crystal lattice, as well as the differences in the orientation distribution of crystalline blocks and participation of other phases. AC magnetic susceptibility magnetometry, performed with Physical Property Measurement System (PPMS, Quantum Design), was applied at various temperatures and at the frequency of 10 kHz, to ascertain the difference in the integrated superconducting properties. Measurements of the field dependence of magnetization were carried out using Magnetic Property Measurement System (MPMS-5, Quantum Design).

Detailed transport measurements of two crystals, well characterized structurally and magnetically, grown with velocities ~ 5 (sample A) and ~ 1 mm/h (sample B), were performed in order to study the local superconducting and transport properties of the samples. The four-probe method of measurement of voltage $V$ at direct current $I$, flowing through the bridges which were cut out from the samples by means of a laser beam [19] and pasted on quartz plates, was utilised. The bridges' characteristic sizes were as follows: the length was in the range of 200–300 μm, the width was in the range of 10–90 μm, and the thickness was in the range of 20–30 μm.

Figure 1(a) shows the photographs of the studied crystals. The quantitative point chemical composition analyses were performed on the natural cleavage planes of the crystals, applying Oxford INCA 250 energy dispersive x-ray spectroscopy (EDX) coupled with the JEOL JSM-7600F field emission (Schottky type) scanning electron microscope (FESEM), operating at 20 kV incident energy. The average chemical composition of the crystals was determined as $Fe_{1.00}Te_{0.66}Se_{0.34}$ and $Fe_{1.01}Te_{0.66}Se_{0.34}$ for samples A and B, respectively. The FESEM images of the (001) crystal planes (figure 1(b)) shows two different parts of sample A (two left photos) and sample B (two right photos). Although, in the case of sample A, some regions exhibit comparable crystalline quality with that of sample B, in general, sample B shows smoother and smaller surface steps and is apparently a good quality single crystal. The x-ray powder diffraction patterns of the powdered crystals were recorded at room temperature using a Siemens D500, equipped with high-resolution Si:Li detector, and DRON3 diffractometers with Ni-filtered Cu $K_\alpha$ radiation (see figures 2(a)-(c)). The diffractograms were analysed by the Rietveld refinement method using the DBWS-9807 program [20].

## 4. Results and discussion

Analysing the x-ray patterns, major tetragonal phase reflections were indexed assuming a tetragonal cell in the space group *P*4/*nmm* (No. 129) of the PbO structural type with occupation Wyckoff's 2*a* site by Fe, and the 2*c* site by Se/Te. Accurate values of the *c* lattice constant and the Δω value – describing the FWHM of the rocking curve ω scan on the 004 diffraction peak – were obtained on the well-defined, natural cleavage (001) plane. The Δω value was chosen as a criterion of the crystallographic quality of the obtained crystals, and it was found to be equal to ~ 6.0' for sample A and ~ 1.4' for sample B. The *c* lattice constant was used as a fixed value in the powder Rietveld analysis for the determination of other structural parameters. The lattice parameters of the major phase are *a* = 3.799 Å, *c* = 6.090 Å for sample A and *a* = 3.799 Å, *c* = 6.093 Å for sample B.

Powder diffraction patterns for powderised crystals A (figure 2(a)) and B (figure 2(b)) show some differences. The diffraction pattern of sample A (figure 2(a)) shows the additional peaks of the Fe deficient $Fe_7(Se,Te)_8$ hexagonal-like phase P3$_1$21 (no.152), marked in figure 2(a) with the symbol "#". In contrast, such peaks are not visible in the diffraction pattern of sample B (figure 2(b)). This may be due to either a very small size (of the order of nanometres) of the hexagonal-like phase crystallites, or the absence of these phases in crystals with a low growth rate. In our previous studies [18], we noted a tendency to organize Fe vacancies into nanometre clusters, leading to the creation of hexagonal-like structure regions in the tetragonal host matrix, with the size and distribution depending on the speed of crystal growth, being undetectable in x-ray measurements. There is also a difference in the diffraction patterns among the intensities of the 00$l$ type peaks, being much better developed for sample B than for sample A. This is due to difficulties with powderisation of the high crystallographic quality crystal (sample B) because the natural cleavage planes, not leading to isotropic grain orientation distribution – texturisation.

The analysis of the x-ray data obtained for the single crystal measurements in the Bragg-Brentano geometry certified that sample B, grown at a small speed of 1 mm/h, is a single crystal of very high crystallographic quality (figure 2(c)). The x-ray patterns received from the mirror reflecting planes of the system of cleaving steps indicate that these planes of the $Fe_{1.01}Te_{0.66}Se_{0.34}$ sample are the basic planes (001). These planes are union planes of the strongly anisotropic tetragonal crystal, as has been previously discussed in several papers [17, 18, 21]. The shape of cleaving surfaces and the received x-ray patterns confirm the one-phase state and high perfection of the crystal grown with small speed. The received x-ray patterns of the crystal contain tetragonal phase reflections only, and exhibit extremely high intensity and small broadening of the 004 diffraction peak. The appearance of weak "peaks" to the left of the basic lines (marked by symbol "β" in figure 2(c)) is a consequence of the insufficient efficiency of the Ni-filter for the absorption of β-radiations at very effective diffraction or superstructure of the investigated crystal.

The diffraction pattern for the poorer crystallographic quality sample A, similar to that one presented in figure 2(c) for sample B, does not provide reliable data. Distribution of the $c$-axis orientation of the sample A blocks is too wide, so the cleavage plane is not well defined. One can find in the diffraction pattern (not presented) that, except for the main tetragonal phase, the areas with a crystal structure of a hexagonal-like phase and, probably, even phases with lower symmetry, are also visible there. A small volume fraction of the crystals with a grain size smaller than 0.1 μm or a small amount of an amorphous phase, concentrated basically in wide inter phase borders, is noticeable as well. The intensity of the diffusion dispersion is large and surpasses the total intensity of the diffraction from the main crystal phases. The intensity of the corresponding reflections 002, 003, and 004 is almost three orders of magnitude smaller than that of sample B (perfect single crystal), despite the fact that the tetragonal phase in such samples has parameters close to the parameters of sample B.

Studies of the transport properties have been carried out using three types of bridges which have been cut out from the crystals. Two types of bridges have been made from sample B and one type was made from sample A. The surfaces of the bridges are shown in figures 3(a), (b), (d). The plane of the first type of bridge made from sample B (figure 3(a)) in the direction of its length and width corresponds to the *a-b* plane of a given crystal (figures 1(a), (b), right side). The bridges made from sample A (figure 3(b)) were cut out from its dense area and have a metal shine in the reflected light. The surface of the second type of bridge, made from sample B (figure 1(b), right side), in the direction of its length and width, was not flat as it was the fused part of the edge side of the crystal (figure 3(d)). The measuring current, *I*, in all bridges has been directed lengthways to their surface. In figure 3(a), the left panel shows the bridge made of the central part of sample B; the white part of the right panel is a scheme of the bridge with a conditional arrangement of current ($I_1$, $I_2$) and voltage ($V_1$, $V_2$) contacts. In figure 3(b), the left panel shows the bridge which was cut off from sample A, and the right panel shows a scheme of the bridge, where one can see four gold wires with a diameter of 50 μm joined with indium contacts, which are rubbed in the crystal surface mechanically. In figure 3(c), the left panel shows sample B from the side of the fused edge in plane *Y-Z*; the right panel shows a scheme of the same sample with an indication (by shaded lines) of the region where the bridge shown in the left panel of figure 3(d) was cut. In figure 3(d), the left panel presents an external view of the bridge made of the fused edge of sample B; the right panel presents a scheme of the same bridge with an indication of the places of the current and voltage contacts.

The structural distinctions of samples A and B influence the specific resistance, ρ, of the bridges. The resistance of the bridges of the first type cut off from sample B determined at $T = 300$ K is equal to $\rho(300\ K) = (6–6.45)\cdot10^{-4}$ Ω·cm and the ratio $\rho(20\ K)/\rho(300\ K) \approx 1.4–1.55$. The bridges of the second type cut off from sample B exhibit $\rho(300\ K) \approx 5.3\cdot10^{-4}$ Ω·cm and $\rho(20\ K)/\rho(300\ K) \approx 1.5$. The bridges which have been cut off from sample A, exhibit $\rho(300\ K) = (2–2.8)\cdot10^{-4}$ Ω·cm and the ratio $\rho(20\ K)/(300\ K) \approx 1.17–1.25$. One can see that the specific resistance of the bridges cut off from sample A is at least twice as small as that of the bridges made from sample B.

It is important to note that the value of $\rho(300\ K) = 6\cdot10^{-4}$ Ω·cm measured for similar crystals of FeTe$_{0.61}$Se$_{0.39}$ [21] corresponds to the values obtained for the bridges of the first type cut off from sample B. However, the resistance of the crystals of FeTe$_{0.61}$Se$_{0.39}$ decreases with decreasing temperature [21], whereas the resistance of the bridges made from samples B and A increases with decreasing temperature.

Importantly, the performed transport measurements allowed for a precise determination of the critical temperature for the studied samples – see discussion in the next paragraph. These measurements have also confirmed an unexpected correlation between the width of the transition and microstructure of the samples having an identical composition. Temperature dependences of a real part of complex magnetic susceptibility for samples A and B, recorded at 10 kHz of excitation frequency at the value of a magnetic field of 1 Oe are shown in figure 4(a). Both samples A and B exhibit almost the same onset of $T_c$ ($T^{onset}_c$ ~ 13.7 K±0.2 K), as determined from the temperature dependence of the real part of the AC susceptibility $4\pi\chi'$, despite the qualitatively difference in the shape of the susceptibility curves and significant width of the transition to the superconducting state. The width of the transition for sample A of apparently worse crystallographic quality is twice as small as that for a practically ideal single crystal (sample B).

The temperature dependence of resistivity for all bridges cut off from both studied samples (figures 3(a)-(c)) shows very similar $T^{onset}_c \approx 16$ K (see figure 4(b), where it was assumed that at $j = 10$ A/cm$^2$ a state of the bridge with $R = 0$ Ω arises). The difference in both resistivity and magnetic $T^{onset}_c$ among the samples does not exceed 1 K, whereas the difference in $T^{onset}_c$ determined from the resistivity and from the magnetic measurements is equal to about 2.3 K. However, the resistivity transition width, $\Delta T_c = T^{onset}_c - T_{c0}$, where $T_{c0}$ is the maximal temperature at which the bridge resistance become negligible, $R = 0$, depends on the type of bridge, but is always bigger for the bridges cut off from sample B (in comparison with the bridges cut off from sample A).

One can also notice that the transport measurements of the transition temperature and critical current performed for the microscopic bridges allow us to define the $T^{onset}_c$ more precisely than those based on magnetic susceptibility measurements, and they allow us to deduce not only the integrated but also the local properties of a superconductor.

The value of $j_c$ at fixed temperature (for example the data at $T \leq 10$ K) of the first type of bridge made on sample B is about 75 times smaller than the value of $j_c$ for the second type of bridge made on the same sample, and more than 1000 times smaller than the value of $j_c$ for the bridge made on sample A.

As one can see, comparing figures 4(a) and (b) with figure 4(c), the small width of the superconducting transition, as a rule, correlates very well with the high critical current density, $j_c$, recorded at temperatures essentially lower than $T^{onset}_c$. On the contrary, for the samples with a large transition width, the critical current density, as a rule, is small.

Considering the possible reasons for the difference in the width of transition and in the critical current density observed for the studied samples, we should take into account that the very similar value of resistivity $T^{onset}_c$ for all bridges proves that all of them consist of an identical superconductor with an almost identical chemical composition. The difference consists mainly in the different ability of the bridges and samples as a whole to carry a superconducting current at temperatures $T < T_{c0}$. This, in turn, is determined by the path a superconducting current takes. For the bulk sample, in principle, two "building block" factors are relevant: large single crystalline regions and links among them. On the other hand, one should take into account that an average Fe concentration differs slightly for both samples (the compositions determined by EDX are: $Fe_{1.00}Te_{0.66}Se_{0.34}$ and $Fe_{1.01}Te_{0.66}Se_{0.34}$ for sample A and B, respectively) and the actual values of excess Fe may deviate to some extent from the average value determined by EDX. Since an excess Fe at the interlayer site greatly affects the superconducting properties, we take into account that this factor may also cause a difference in the critical current properties between samples A and B.

First we shall consider sample B. The planes of the micro single crystals of that sample are parallel each other and their thickness is equal to about 30 μm. Along a length of the bridge on its surface, some borders of the micro single crystals in the form of the "steps" are visible (figure 3(a)). These steps are the reason for the heterogeneity of the thickness of the bridge along its length. The width of the transition to the superconducting state, as well as the critical current density, is determined by the structure and properties of the links between the micro single crystals. It is known that layers of the chalcogenides are poorly mechanically connected with each other. For example, there is a technique of manual stratification of such single crystals by means of an estrangement of the parts of the layers by an adhesive tape. Casual imperfections can be introduced between the layers of a crystal in its growth process. As a result, the set of so-called weak superconducting links between the layers may be formed. A set of tunnel Josephson contacts with a layer of an insulator (S-I-S contact) and/or of normal metal (S-N-S contact) may be developed. One should take into account that at a higher growth rate, imperfections remain at the weak links and act as pinning centers together with the small amounts of secondary phases. Additionally, in the sample grown at faster rate, the weak links seem to be more metallic and crystallite connectivity seems to be better (as indicated by the lower resistivity and higher $j_c$). The character of weak links – metallic or insulating – strongly affects the $j_c(T)$ dependence and resistivity due to the proximity effect and Cooper pair tunneling. The value of $j_c$ of these contacts strongly depends on temperature, and also on the thickness of an insulator and normal metal in the contacts. As a consequence, the critical current density of the links can vary in a wide range of values, both along the length of the bridge and from one layer up to another deep into the thickness of the bridge. The significant distinction of the critical currents of the weak links between the layers is assumed to be the reason for the wide superconducting transition.

Now we shall consider the properties of sample A. A mechanical test of its surface indicates essentially greater heterogeneity in comparison with those of sample B. Friable regions with a small density alternate with more dense regions which prevail. X-ray studies confirm the existence of variously orientated micro single crystals and the presence of other phases. The bridges made of a dense part of this sample have lower specific resistance than the bridges of sample B. Thus it is necessary to note that the growth velocity of sample A was five times larger than the growth velocity of sample B. The weak links between the micro single crystals of the dense part of the sample are of better quality. They exist not only between the flat parts of the micro single crystals, but also between the end faces of the micro single crystals. Such weak links usually are of an S-N-S type or are metal micro bridges between the micro single crystals. These contacts have great values of $j_c$ and exhibit a smaller distribution of these values from one contact to other. In consequence, one observes a narrower superconducting transition and essentially higher $j_c$ values for sample A.

The bridges of the second type cut off from sample B have been made of the fused part of the sample edge. The properties of these bridges are essentially different from the properties of the bridges made from the other, non-fused part of the same sample. These properties are closer to the properties of the bridges cut off from sample A. This indicates that the properties of the weak superconducting links in these bridges are similar to the properties of the links in the bridges made from sample A. Unfortunately, the reason for the formation of the specified fused edge of sample B is unknown.

Magnetic measurements performed for high magnetic fields allowed us to determine the position of the irreversibility line (IL), values of the critical current density estimated from the hysteresis loop width and to check the scaling of the pinning force. As one can see in figure 5(a), the position of the IL is quite similar for both samples. The dependence of the irreversibility field $H_{irr}$ on temperature:

$$H_{irr}(T) = H_{irr}(0)(1 - T/T_c)^n \tag{1}$$

describes well the position of the IL for the same index $n= 2$. Such a value of $n$ was found a long time ago in bismuth-based high-$T_c$ cuprates (see, e.g. [22]). These compounds exhibit a strong layered structure. The data presented in figure 5(b) confirm that the values of $j_c$ for sample A are at least one order of magnitude larger than those for sample B. The values of $j_c$ were determined from the width of the hysteresis loop, assuming that the current circulate around the loop with a diameter of the whole studied sample equal to about 2.4 mm for both samples. Due to the uncertainty in the determination of these diameters, the absolute values of $j_c$ are estimated roughly only, and in the case of sample A, the estimated values are rather the lower limit of $j_c$ since, due to the structure (crystallinity) of this sample, the current may circulate around a smaller loop. The reduced pinning force $F_p/F_{pmax}$ versus $H/H_{irr}$ is shown in figure 5(c). One can see that the pinning force for sample B scales with temperature, which is not the case for sample A. Sample A contains various inclusions of other phases (defects) that determine the pinning. Due to the spread in dimensions of these defects, the pinning force does not scale. It is consistent with the difference in $j_c(T)$ dependence noticed in the transport measurement for the bridges made from samples A and B (inset to figure 4(c)).

## 5. Conclusions

The studies of two superconducting $FeTe_{0.65}Se_{0.35}$ samples have shown that the growing conditions of these crystals essentially influence their superconducting transport properties. The resistivity onset temperature of superconducting transition $T^{onset}_c$ of these samples is very similar, which indicates the stoichiometric composition of both samples.

The almost ideal single crystal of $Fe_{1.01}Te_{0.66}Se_{0.34}$ exhibits a greater width of superconducting transition and a considerably smaller value of critical current density in comparison with the non-uniform sample of the same compound. The main difference in the growing process of these samples consists in five times slower growth velocity of an ideal crystal in comparison with the non-uniform one. The most probable reason for the difference in the superconducting properties of the samples is the different period of formation of the investigated samples. For the shorter period of formation (sample A), the imperfection remains in the layers of the micro single crystals, forming the additional pinning centers for a superconducting current. Moreover, in the sample grown at a faster rate, the weak links seem to be more metallic, and the crystallite connectivity seems to be better (lower resistivity, higher $j_c$). Metallicity versus insulating character of the weak links strongly affects $j_c(T)$ dependence and resistivity, via the proximity effect and Cooper pair tunneling. This improves the superconducting transport properties of the given sample and leads to higher value of critical current density. On the other hand, the results confirm that the inhomogeneous spatial distribution of ions and small hexagonal-like phase chalcogenides with nanoscale phase separation seems to enhance the superconductivity in this system. The conclusions drawn from magnetic measurements are in line with those drawn from transport measurements.

**Acknowledgments**

This work was supported by the National Science Centre of Poland based on decision No. DEC-2013/08/M/ST3/00927. We would like to thank M. Kozłowski for experimental support.

**Figures**

**Figure 1.** (a) Photography of sample A (left panel) and sample B (right panel) with exhibited natural (001) crystallographic planes. The grid step corresponds to 1 mm. (b) FESEM images of the (001) crystal planes for samples A (left panels) and B (right panels).

**Figure 2.** (a) Powder x-ray diffraction pattern, with reflections indexed assuming a tetragonal cell in the space group *P*4/*nmm*, of: (a) sample A (the additional peaks of the hexagonal-like $Fe_7(Se,Te)_8$ phase are marked with the symbol "#"), (b) sample B. (c) The x-ray diffraction pattern from planes (00l) of the sample B, received with Bragg-Brentano geometry on diffractometer DRON-3 with Cu $K_\alpha$ radiation. The corresponding indexes of the planes of reflection for a tetragonal phase are specified.

**Figure 3.** Micro photos: (a) The left panel shows the bridge made of the central part of sample B; the white part of the right panel is a scheme of the bridge. (b) The left panel shows the bridge which was cut off from sample A; the right panel is a scheme of the bridge. (c) The left panel shows sample B from side of the fused edge in plane *Y-Z*; the right panel presents a scheme of the same sample, with an indication (by shaded lines) of the region where the bridge was cut shown in the left panel of figure 3(d). (d) The left panel shows external view of the bridge made of the fused edge of the fused edge of sample B; the right panel presents a scheme of the same bridge.

**Figure 4.** (a) Temperature dependences of the real part ($4\pi\chi'$ – lower panel) and the imaginary part ($4\pi\chi''$ – upper panel) of the AC magnetic susceptibility, normalized to the ideal value of -1 for the real part of the AC susceptibility, measured in 1 Oe of AC field with 10 kHz in warming mode for two $FeTe_{0.65}Se_{0.35}$ single crystals of significantly different crystallographic qualities. The presented data were normalized for better comparison of the susceptibility data obtained for the samples with different shapes and, therefore, subjected to different demagnetizing fields. (b) The temperature dependences of the resistivity of different bridges made from the light part of sample A, the fused edge of sample B, and the middle part of sample B. The inset shows temperature dependence of resistivity in an extended temperature range. (c) The temperature dependences of $j_c$ measured for the same bridges. The inset shows $j_c(T)$ dependence with a semi-logarithmic scale.

**Figure 5.** (a) Irreversibility line for samples A and B presented in double logarithmic scale. (b) Field dependence of the critical current density, determined from the width of the hysteresis loop, for samples A and B. (c) Normalized pinning force dependence on $H/H_{irr}$ for samples A and B.

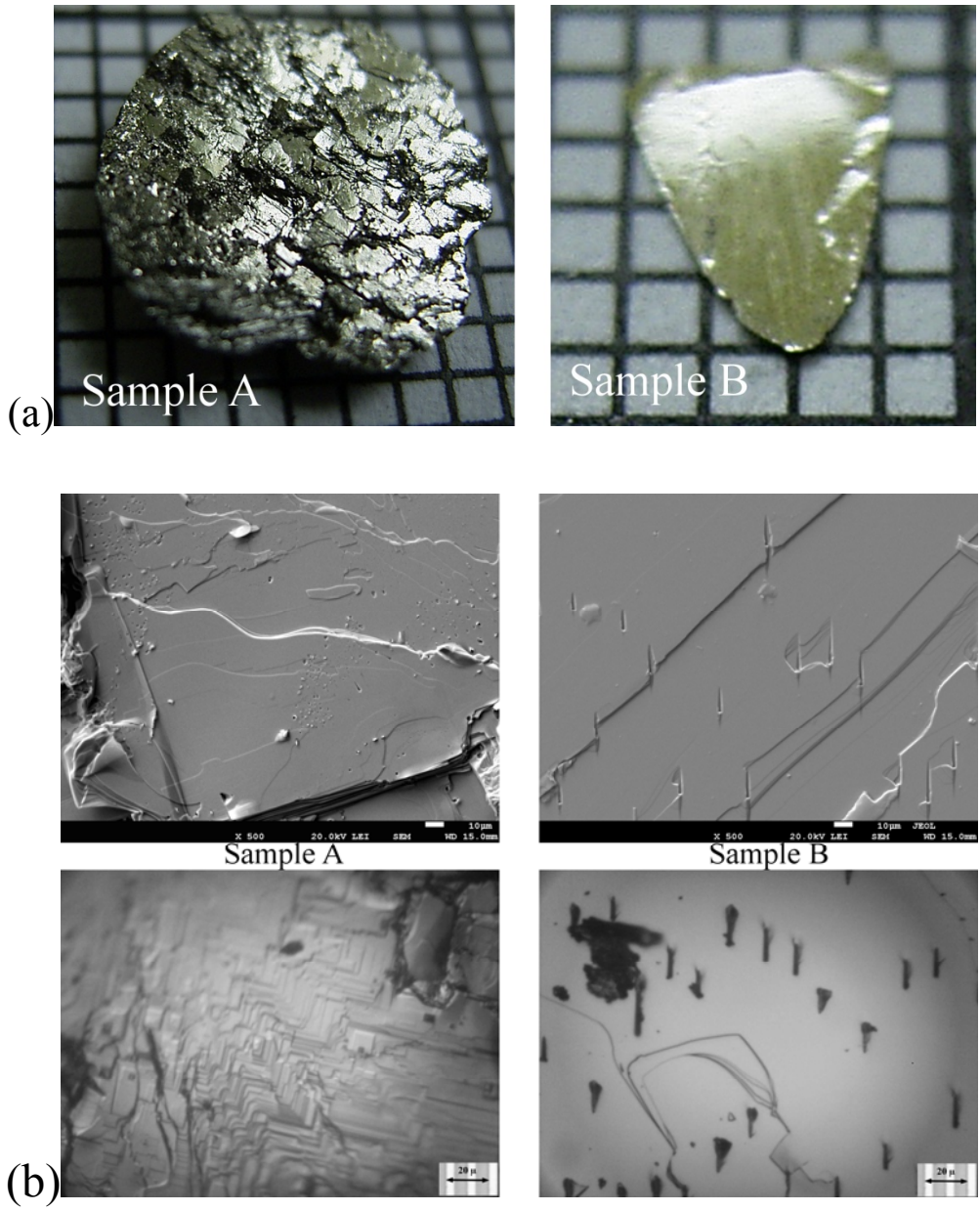

**Figure 1**

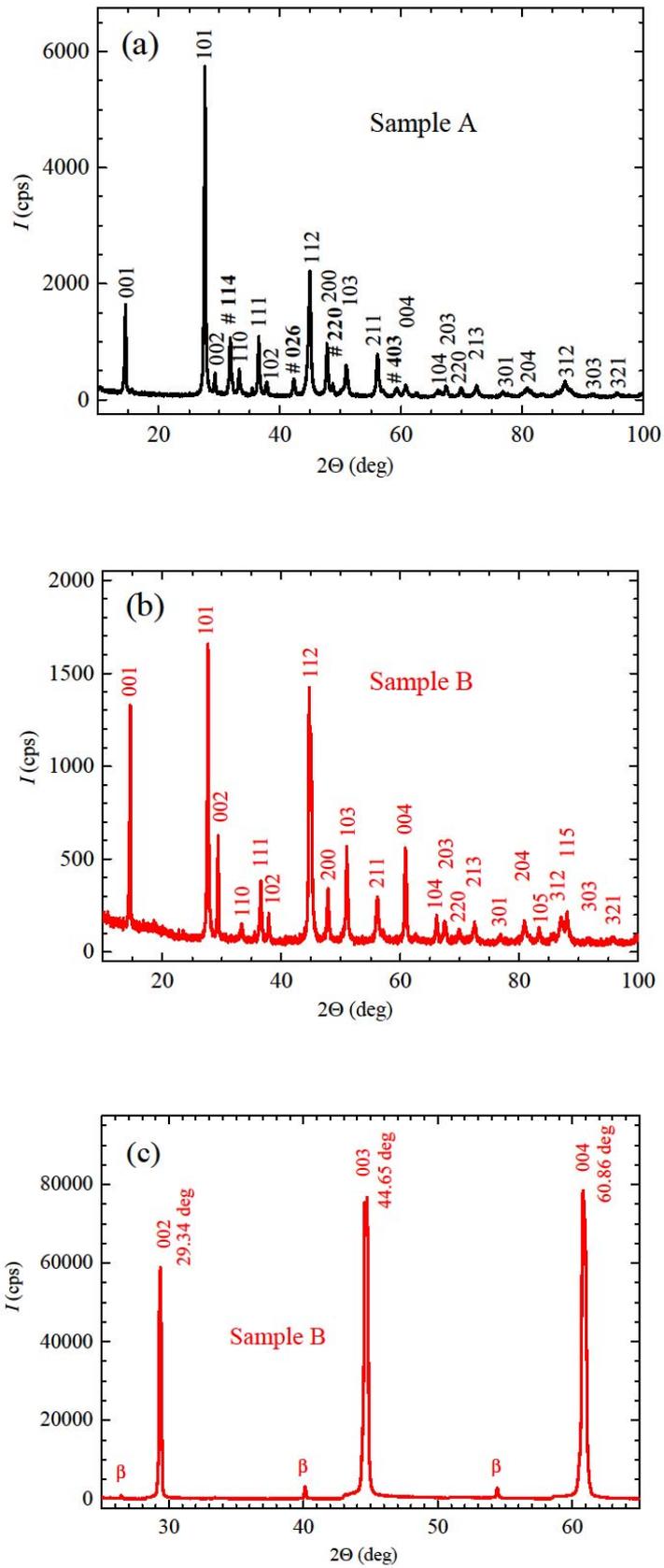

**Figure 2**

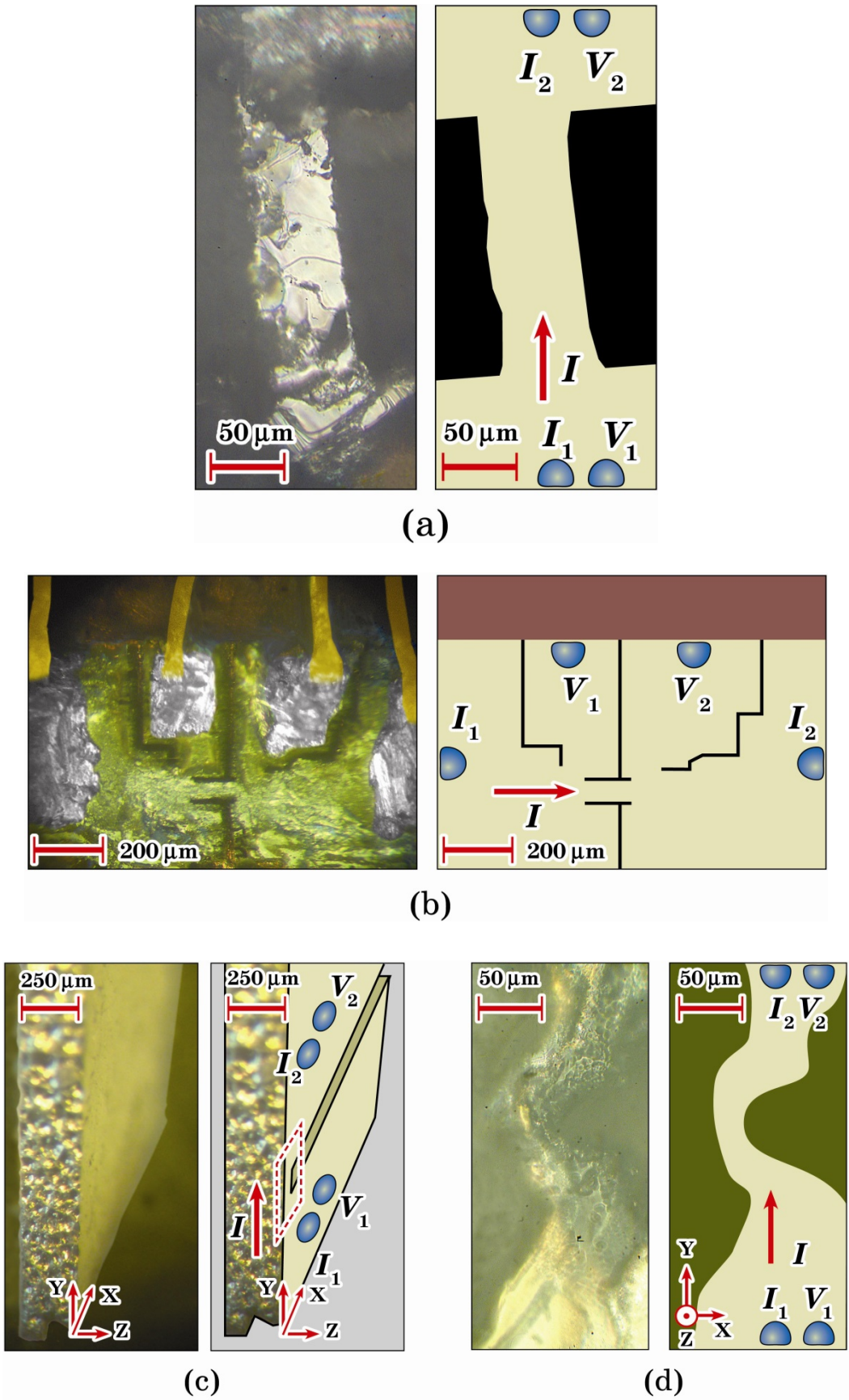

**Figure 3**

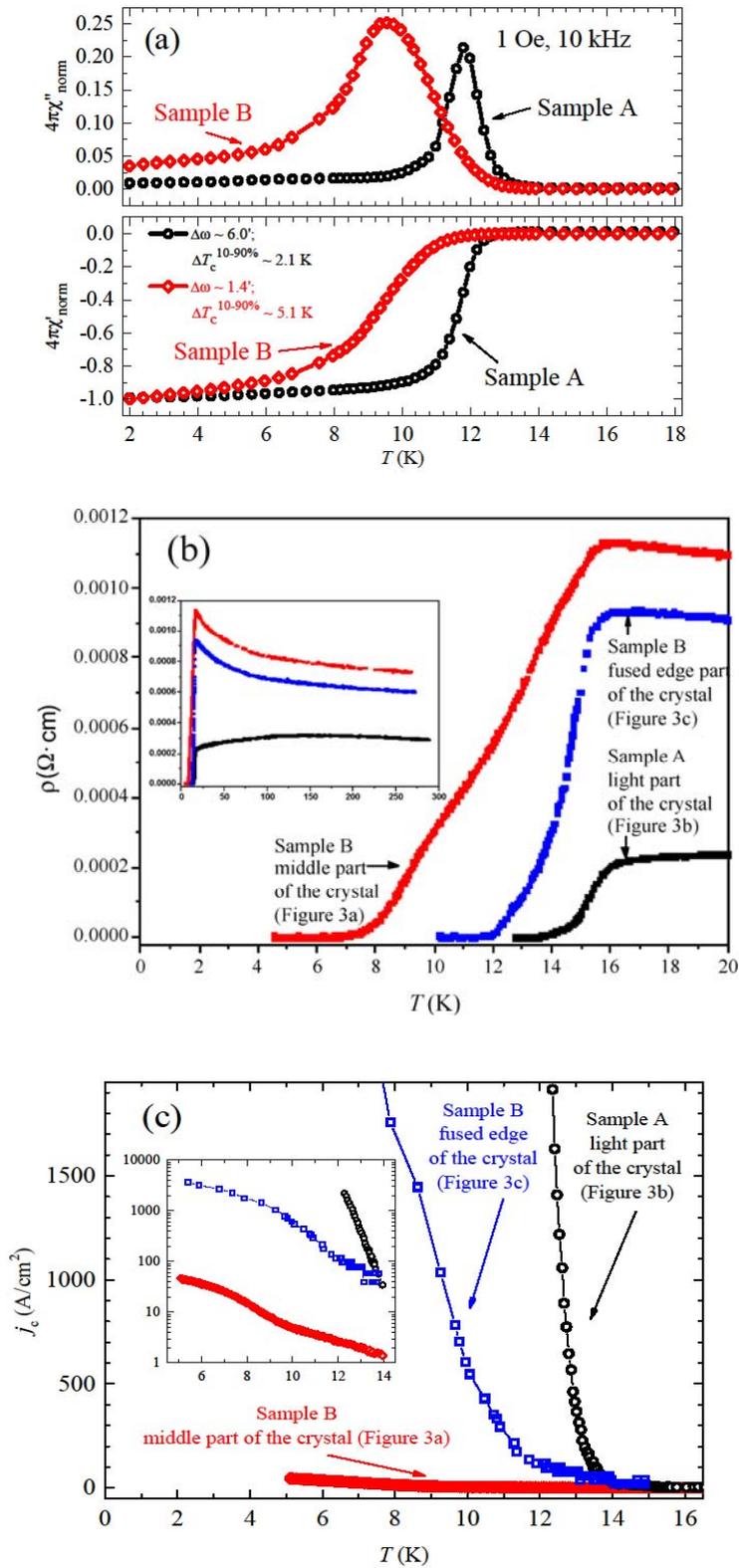

**Figure 4**

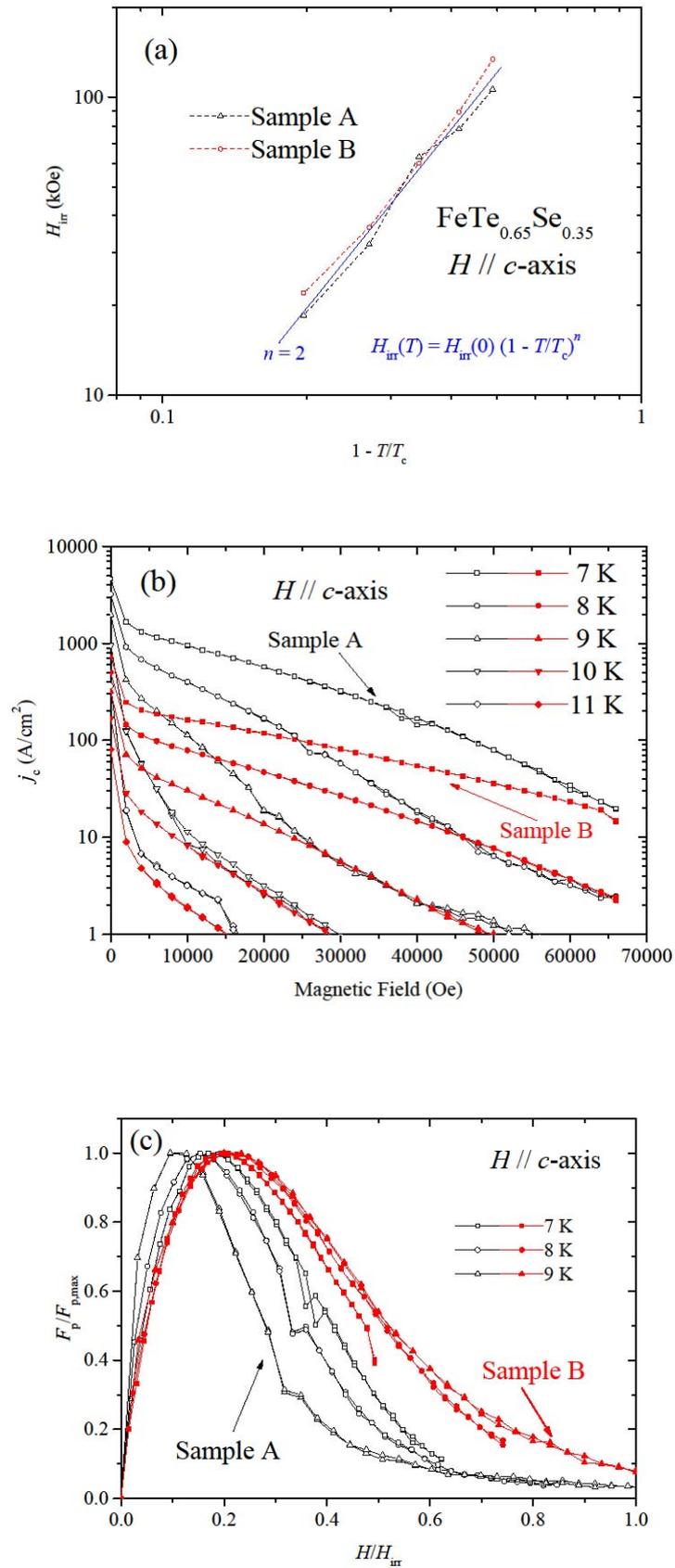

**Figure 5**